\documentclass[12pt]{article}
\newcommand{\be}{\begin{equation}}
\newcommand{\ee}{\end{equation}}
\newcommand{\bea}{\begin{eqnarray}}
\newcommand{\eea}{\end{eqnarray}}
\newcommand{\sptwo}{1.4}
\newcommand{\doublespace}{\edef\baselinestretch{\sptwo}\Large\normalsize}
\newcommand{\newsection}[1]{\section{#1}\setcounter{equation}{0}}

\newcounter{newapp}
\begin{document}
\begin{titlepage}
\vspace*{1.0in}
\begin{center}
{\large\bf The Effective Potential And Additional \\
Large Radius Compactified Space-Time Dimensions}
\end{center}
\vspace{0.25in}
~\\
\begin{center}
{\bf T.E. Clark\footnote{e-mail address: clark@physics.purdue.edu} and S.T. Love\footnote{e-mail address: love@physics.purdue.edu}}\\
{\it Department of Physics\\ 
Purdue University\\
West Lafayette, IN 47907-1396}
~\\
~\\
\end{center}
\vspace{1in}
\begin{center}
{\bf Abstract}
\end{center}

The consequences of large radius extra space-time compactified dimensions on the four dimensional one loop effective potential are investigated for a model which includes  scalar self interactions and Yukawa coupling to fermions.  The Kaluza-Klein tower of states associated with the extra compact dimensions shifts the location of the effective potential minimum and modifies its curvature. The dependence of these effects on the radius of the extra dimension is illustrated for various choices of coupling constants and masses. For  large radii, the consequence of twisting the fermion boundary condition on the compactified dimensions is numerically found to produce but a negligible effect on the effective potential.
\end{titlepage}
\pagebreak

\doublespace

\newsection{Introduction}

A common feature for the consistent formulation of many currently proposed fundamental theories is the necessity of more than four space-time dimensions. Since these extra dimensions are presently unobserved, they are assumed to be compactified.  Associated with the compactified dimensions, there appears a tower of Kaluza-Klein (KK) field excitations in four space-time dimensions whose masses are integer multiples of the inverse compactification radius $R$.  In models where $R$ is the order of the Planck length ($\sim 10^{-33}$ cm ) or smaller, the superheavy KK modes inconsequentially decouple from the low energy physics.  On the other hand, for radii much larger than the Planck length, low energy physics can be considerably affected by the propagation of fields into the large radii extra dimensions.  Since theories formulated in higher dimensions tend to have a more divergent short distance behavior which leads to the breakdown of renormalizability, an ultraviolet cutoff $\Lambda$ must be included for a consistent interpretation of such field theories.  For large compactification radii, $\Lambda R >1$, the KK modes with mass below the cutoff could produce various interesting theoretical and phenomenological consequences \cite{SSCSS}-\cite{CLK}.  

Many analyses of the consequences of the higher dimensions have focused on their ramifications on the minimal supersymmetric standard model or one of its extensions  \cite{TR2}-\cite{DDG}. This is quite reasonable since one generally views the higher dimensional underlying theory operating at the very short distance scale to have supersymmetry as one of its vital components.  One could, however, envision a scenario where the supersymmetry breaking scale is larger than the inverse radius of the compactified dimensions in which case one should more appropriately study the effects of the extra dimensions on the standard model itself.  For such a relationship between scales, as one scales down in energy, it is the SUSY partners which decouple first (at a higher energy) before the tower of Kaluza-Klein states associated with the standard model degrees of freedom.

In this note, we explore the ramifications of just such a scenario.  We focus attention on the effects of the Kaluza-Klein tower of states coming from a compactified fifth dimension on the one loop effective potential for a simplified model consisting of a self-coupled real scalar field and a Yukawa coupled spinor field.  In five dimensions,  the fermion field is taken as an eight component, six dimensional Dirac spinor, which reduces to a pair of towers of four component Dirac spinors in four dimensions.  In particular, we focus on the modifications to the one loop effective potential for the zero mode four dimensional scalar field when the product of the ultraviolet cutoff, $\Lambda$, and the compactification radius, $R$, is large, $\Lambda R > 1$.  The dependence of the effective potential on the radius of compactification, $R$, will be illustrated for typical values of the tree level coupling constants.  The radius dependences of the vacuum value of the scalar field and the effective potential curvature at this minimum are determined.  Finally, the form of the effective potential is shown to be quite insensitive to the particular boundary conditions imposed on the fermion fields for large compactification radii.

\newsection{The Effective Potential}

Denoting the usual four dimensional space-time coordinates of  Minkowski space by $x^\mu$, while $0\leq y\leq 2\pi R$ is the coordinate of the compact fifth dimension, the action for the self-coupled hermitian scalar field, $\phi (x,y)$ and eight component Dirac fermion, $\Psi (x,y)$, is given by
\bea
I &=& \int d^4x \int_0^{2\pi R} dy \left\{ \frac{1}{2}\partial_M \phi\partial^M\phi -V^{(5)}(\phi) \right.\cr
 & &\left. +\bar\Psi \gamma^M i\partial_M \Psi -M_f \bar\Psi\Psi -g^{(5)}(\phi)\phi\bar\Psi\Psi 
\right\}.
\eea
Here $V^{(5)}(\phi)$ is the scalar field potential describing its self-interactions, $g^{(5)}(\phi)$ is a scalar field dependent generalized Yukawa coupling and $M_f$ is an explicit fermion mass.  The five $\gamma^M$, with $M=0, 1, 2, 3, 4$, are five $8\times 8$ Dirac gamma matrices. Imposing untwisted boundary conditions on the scalar field, $\phi (x,y+2\pi R) = \phi (x,y)$, it may be expanded in a Fourier series as
\bea
\phi (x,y) &=& \frac{1}{\sqrt{2\pi R}}\sum_{n=-\infty}^{+\infty} \phi_n(x) e^{in\frac{y}{R}}\cr
 &=& \frac{1}{\sqrt{2\pi R}} \left\{\phi_0 (x) +2 \sum_{n=1}^{\infty} \phi_n(x) \cos{(n\frac{y}{R})}\right\} ,
\label{FS1}
\eea
where the second equality follows as a consequence of the hermiticity of the field, which dictates that $\phi_{-n}=\phi_n$.  On the other hand, the Dirac field is allowed to assume arbitrarily twisted boundary conditions so that $\Psi (x, y+2\pi R)=e^{2\pi i a}\Psi(x, y)$, with $a$ an arbitrary real number.  Thus the fermion field Fourier expansion is given by
\be
\Psi (x,y) =\frac{1}{\sqrt{2\pi R}}\sum_{n=-\infty}^{\infty} \psi_n (x) e^{i (n+a)\frac{y}{R}} .
\label{FS2}
\ee

The one-loop contribution to the effective potential, $V_{\rm eff}(\varphi)$, for the zero mode scalar field $\phi_0(x)=\varphi(x)$ is found by summing over all one loop graphs with zero momentum, zero mode external $\varphi$ lines, but with propagating internal lines consisting of all zero modes, $\phi_0$ and $\psi_0$, and all KK tower modes, $\phi_n$ and $\psi_n$, for $n\neq 0$. To obtain such graphs, one need expand the action $I$ about $\phi_0=\varphi$ and retain terms quadratic in the $\phi_n$ and $\psi_n$ fields for all $n$.  Substituting the Fourier series (\ref{FS1}) and (\ref{FS2}) into the five dimensional action and performing the expansion yields an equivalent four dimensional action containing an infinite tower of Kaluza-Klein fields, differentiated from each other only by their masses which arise from the fields' momentum in the direction of the fifth dimension.  The action through second order becomes
\bea
\label{IQUAD}
I_{\rm quad} &=&-\int d^4 x V(\varphi) -\int d^4x \sum_{n=-\infty}^{\infty} \left\{ \frac{1}{2}\phi_{-n}
\left[ \partial^2 +V^{\prime\prime}(\varphi) + \frac{n^2}{R^2}\right]\phi_{n}\right.\cr
 & &\left.+\bar\psi_n \left[ -\gamma^\mu i\partial_\mu +M_f  +g(\varphi)\varphi +\frac{n}{R}\gamma^4 \right] \psi_n \right\} ,
\eea
where conservation of momentum in the direction of the fifth dimension has been used.  The four dimensional potential, $V$, and generalized Yukawa coupling, $g$, are obtained from the five dimension versions $V^{(5)}$ and $g^{(5)}$ by retaining only the zero mode fields and rescaling all couplings to their canonical four dimensional scale by appropriate powers of the size of the fifth dimension.  Effectively, this amounts to replacing $\phi$ by $\varphi$, all couplings by their four dimensional version and accounting for an overall scale factor of the size, $2\pi R$, of the compactified dimension.  We employ the notation where the primes denote differentiation with respect to $\varphi$ so that, for example, $V^\prime \equiv \frac{dV}{d\varphi}$. Using the action ($\ref{IQUAD}$), the 1-loop graphs can be evaluated using standard techniques. After rotating to Euclidean space and performing the angular integrals and the eight dimensional Dirac trace, the one loop effective action for the zero mode scalar field $\varphi$ is 
\bea
V_{\rm eff} (\varphi) &=& V(\varphi)-V(\mu)+\frac{1}{32\pi^2}\sum_{n=-\infty}^{+\infty}\int_0^\infty d\xi \xi \ln{\left[\frac{\xi +V^{\prime\prime} (\varphi)+\frac{n^2}{R^2}}{\xi +V^{\prime\prime} (\mu)+\frac{n^2}{R^2}}\right]}\cr
 & & \cr
 & &-\frac{8}{32\pi^2}\sum_{n=-\infty}^{+\infty}\int_0^\infty d\xi \xi \ln{\left[\frac{\xi +(M_f+g(\varphi)\varphi)^2 +\frac{(n+a)^2}{R^2}}{\xi +(M_f+g(\mu)\mu)^2 +\frac{(n+a)^2}{R^2}}\right]},
\eea
where $\mu$ is an arbitrary normalization scale. 
Introducing a short distance regulated heat kernal representation for the logarithms,
\be
\ln{A}=-\int_{\frac{1}{\Lambda^2}}^{\infty} \frac{ds}{s} e^{-sA} ,
\ee
and performing the momentum integrals then yields
\bea
V_{\rm eff}(\varphi) &=& V(\varphi)-V(\mu)-\frac{1}{32\pi^2} \int_{\frac{1}{\Lambda^2}}^{\infty}  \frac{ds}{s^3}\left[e^{-sV^{\prime\prime}(\varphi)} -e^{-sV^{\prime\prime}(\mu)} \right]\Theta_3(0,e^{-\frac{s}{R^2}}) \cr
 & & \cr
 & &+\frac{8}{32\pi^2} \int_{\frac{1}{\Lambda^2}}^\infty  \frac{ds}{s^3}\left[e^{-s(M_f +g(\varphi)\varphi)^2} -e^{-s(M_f +g(\mu)\mu)^2} \right] e^{-\frac{s}{R^2}a^2} \Theta_3(\frac{isa}{R^2},e^{-\frac{s}{R^2}}) ,\cr
 & & 
\label{Veff1}
\eea
where the Jacobi theta function is defined as
\be
\Theta_3 (\tau, q) = \sum_{n=-\infty}^{\infty} q^{n^2}e^{2in\tau} .
\ee
Note that in the $R\rightarrow 0$ limit, only the zero mode term survives, and the familiar one loop, four dimensional result for the effective potential is reproduced.  For large compact dimensions, $\Lambda R>1$, contributions arise from all modes, and it thus proves useful to employ the Poisson resummation formula
\be
\sum_{n=-\infty}^{\infty}e^{-\frac{s}{R^2}(n+a)^2} =\frac{\sqrt{\pi}R}{\sqrt{s}}\sum_{n=-\infty}^{\infty}e^{-\frac{\pi^2 R^2}{s}n^2} e^{2\pi ina}
\ee
or equivalently
\be
\label{PRF}
e^{-\frac{s}{R^2}a^2}\Theta_3 (\frac{isa}{R^2},e^{-\frac{s}{R^2}}) = \frac{\sqrt{\pi}R}{\sqrt{s}} \Theta_3 (\pi a, e^{-\frac{\pi^2 R^2}{s}}) ,
\ee
to rewrite equation (\ref{Veff1}) as
\bea
V_{\rm eff}(\varphi) &=& V(\varphi)-V(\mu)-\frac{2\pi R}{32\pi^{\frac{5}{2}}} \int_{\frac{1}{\Lambda^2}}^{\infty}  \frac{ds}{s^{\frac{7}{2}}}\left[e^{-sV^{\prime\prime}(\varphi)} -e^{-sV^{\prime\prime}(\mu)} 
\right]\Theta_3(0,e^{-\frac{\pi^2 R^2}{s}}) \cr
 & & \cr
 & &+8\frac{2\pi R}{32\pi^{\frac{5}{2}}}  \int_{\frac{1}{\Lambda^2}}^\infty  \frac{ds}{s^{\frac{7}{2}}}\left[e^{-s(M_f +g(\varphi)\varphi)^2} - e^{-s(M_f +g(\mu)\mu)^2} 
\right] e^{-\frac{s}{R^2}a^2} \Theta_3(\pi a,e^{-\frac{\pi^2 R^2}{s}}) .\cr
 & & 
\label{Veff2}
\eea
The utility of this form for the effective potential becomes apparent when the fifth dimension is large, $\Lambda R \geq 1$. Note that in the $R\rightarrow \infty$ limit, only the zero mode term survives and, with appropriate reversal of the $2\pi R$ dimensional rescaling of the couplings, fields and potential, leads to the familiar result for the five dimensional effective potental 
\bea
V^{(5)}_{\rm eff}(\varphi) &=& V^{(5)}(\varphi)-V^{(5)}(\mu)-\frac{1}{32\pi^{\frac{5}{2}}} \int_{\frac{1}{\Lambda^2}}^{\infty}  \frac{ds}{s^{\frac{7}{2}}}\left[e^{-sV^{(5)\prime\prime}(\varphi)} - e^{-sV^{(5)\prime\prime}(\mu)} 
\right] \cr
 & & \cr
 & &+\frac{8}{32\pi^{\frac{5}{2}}}  \int_{\frac{1}{\Lambda^2}}^\infty  \frac{ds}{s^{\frac{7}{2}}}\left[e^{-s(M_f +g^{(5)}(\varphi)\varphi)^2} - 
e^{-s(M_f +g^{(5)}(\mu)\mu)^2} 
 \right]  .
\eea

In order to facilitate further analysis of the effective potential, it is convenient to 
\begin{figure}[ht]
\vspace*{4.0in}
\special{eps: d: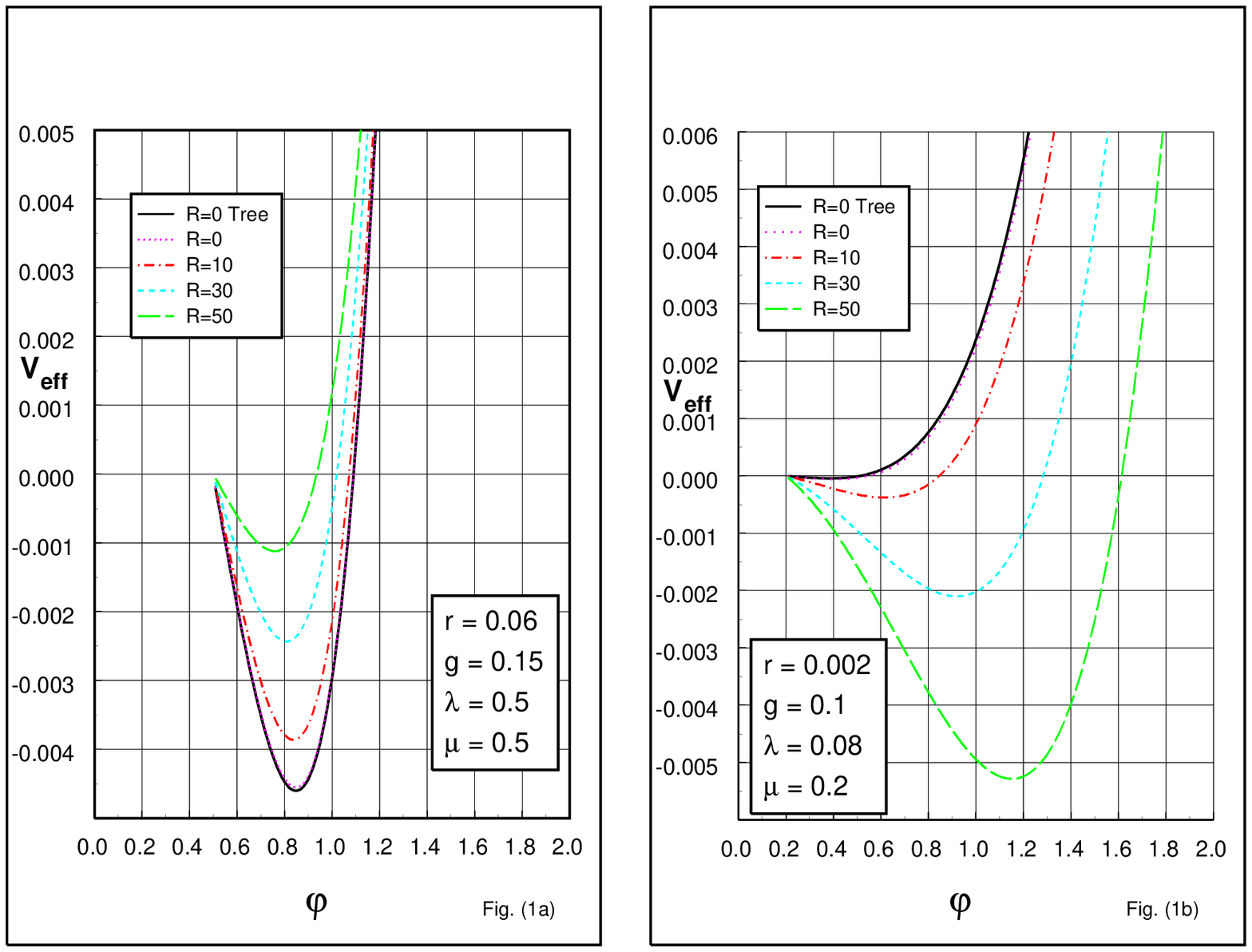 x=6.0in y=4in}
\caption{The effective potential for various sizes $R$ of the compact dimension. In figure (a), the effective potential is normalized so that it vanishes at $\varphi=\mu=.5$, while in figure (b), it is normalized so as to vanish at $\varphi=\mu=.2$. }
\end{figure}
rescale all dimensionful quantities in equation (\ref{Veff2}) by the appropriate powers of the cutoff $\Lambda$ to render them dimensionless. Thus, for example, we let $s\rightarrow s/\Lambda^2$, $\varphi \rightarrow \Lambda \varphi$, $R\rightarrow R/\Lambda$ and $V_{\rm eff}\rightarrow \Lambda^4 V_{\rm eff}$.  With these substitutions the dimensionless effective potential in terms of the now dimensionless zero mode scalar field $\varphi$ and dimensionless radius of compactification $R$, as well as dimensionless couplings and masses, is given by
\bea
V_{\rm eff}(\varphi) &=& V(\varphi)-V(\mu)-\frac{2\pi R}{32\pi^{\frac{5}{2}}} \int_{1}^{\infty}  \frac{ds}{s^{\frac{7}{2}}}\left[e^{-sV^{\prime\prime}(\varphi)} - e^{-sV^{\prime\prime}(\mu)} 
\right]\Theta_3(0,e^{-\frac{\pi^2 R^2}{s}}) \cr
 & & \cr
 & &+8\frac{ 2\pi R}{32\pi^{\frac{5}{2}}}  \int_{1}^\infty  \frac{ds}{s^{\frac{7}{2}}}\left[e^{-s(M_f +g(\varphi)\varphi)^2} -
e^{-s(M_f +g(\mu)\mu)^2} 
 \right] e^{-\frac{s}{R^2}a^2} \Theta_3(\pi a,e^{-\frac{\pi^2 R^2}{s}}) .\cr
 & &
\label{Veff3}
\eea

The explicit factors of $2\pi R$ enhance the radiative corrections to the effective potential as illustrated in Figure 1 where  
the effective potential is plotted versus $\varphi$ for various sizes of the extra compact dimension.  For concreteness, the fermions are taken to have untwisted boundary conditions, $a=0$, no explicit mass, $M_f =0$, and interact with a  constant Yukawa coupling, $g(\varphi)=g$, while the tree scalar potential is taken as
\begin{figure}[ht]
\vspace*{3.0in}
\special{eps: d: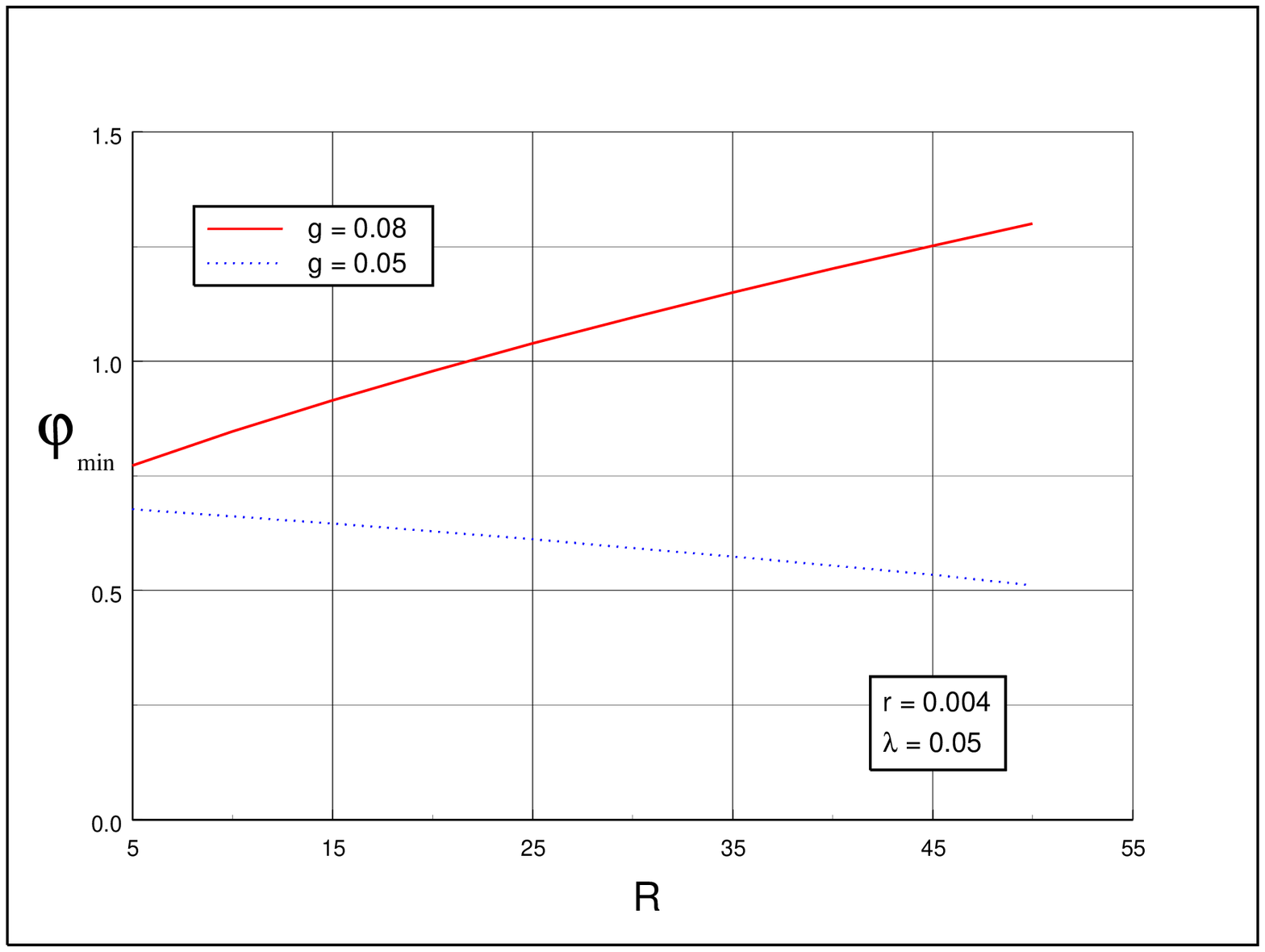 x=6.0in y=3.0in}
\caption{The location of the minimum of the effective potential as a function of the size of the compact dimension $R$ for 2 different values of the Yukawa coupling constant, $g$, with the scalar self coupling constant, $\lambda$ and the dimensionless mass parameter, $r$, held fixed.}
\end{figure}
\be
V(\varphi) = -r\varphi^2 +\frac{\lambda}{4!}\varphi^4 ,
\ee
where the dimensionless mass squared parameter $r$ and the scalar self coupling constant $\lambda$ are chosen so that the potential has a non-trivial minimum. Moreover, the parameters are such that the 1-loop radiative corrections in four dimensions ($R=0$) are quite small as evidenced by the near coincidence of the $R=0$ curves which result from (i) including only the tree potential and (ii) which also includes the 1-loop corrections. In figure 1a, the parameters are chosen such that the radiative corrections are dominated by those graphs with the scalar fields traversing the internal loop (i.e. the scalar self coupling dominates). In this case, as $R$ increases, the minimum of the effective potential tends to decrease while the potential becomes shallower.  On the other hand, when the leading radiative corrections arise from the fermions in the loop (i.e. the Yukawa coupling dominates), as is the case in figure 1b, the location of the minimum increases as $R$ increases, as does the effective potential curvature at its minimum.  This difference in behavior can be directly traced to the relative sign in the radiative correction between the contributions of the scalar and fermion propagating in the loop.

\begin{figure}[ht]
\vspace*{4.0in}
\special{eps: d: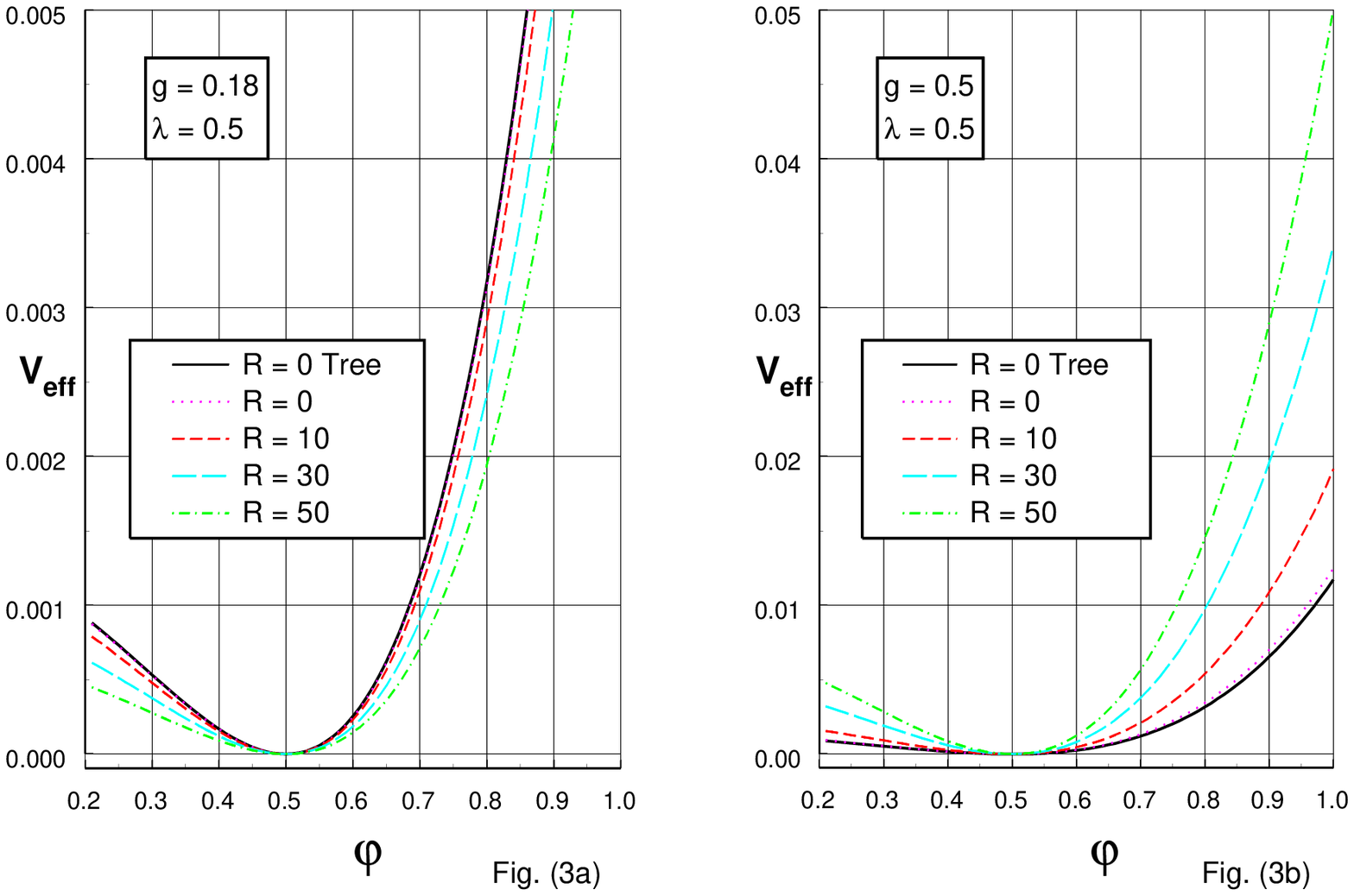 x=6.0in y=4in}
\caption{The effective potential as the size of the compact dimension $R$ varies with the minimum fixed at $\varphi = \varphi_{\rm min} = 0.5$.}
\end{figure}
The location of the effective potential minimum is plotted in figure 2 as a function of $R$ for the two generic cases discussed above. When the radiative corrections are dominated by the scalar self coupling, $\varphi_{\rm min}$ decreases with $R$, while the case of the dominate Yukawa coupling leads to $\varphi_{\rm min}$ increasing with $R$. 
The shape of the curves is essentially determined by the explicit $\sqrt{R}$ dependence of $\varphi_{\rm min}$ which arises from the radiative corrections. For a large extra dimension, $R\geq 10$, the heat kernel is dominated by its ultraviolet contributions (small $s$) where the Poisson resummed Jacobi theta function is very accurately approximated by 1.  Hence the radiative corrections to the effective potential are essentially given as the product of the size of the fifth dimension, $2\pi R$, and the 5 dimensional result. 
\begin{figure}[ht]
\vspace*{3.0in}
\special{eps: d: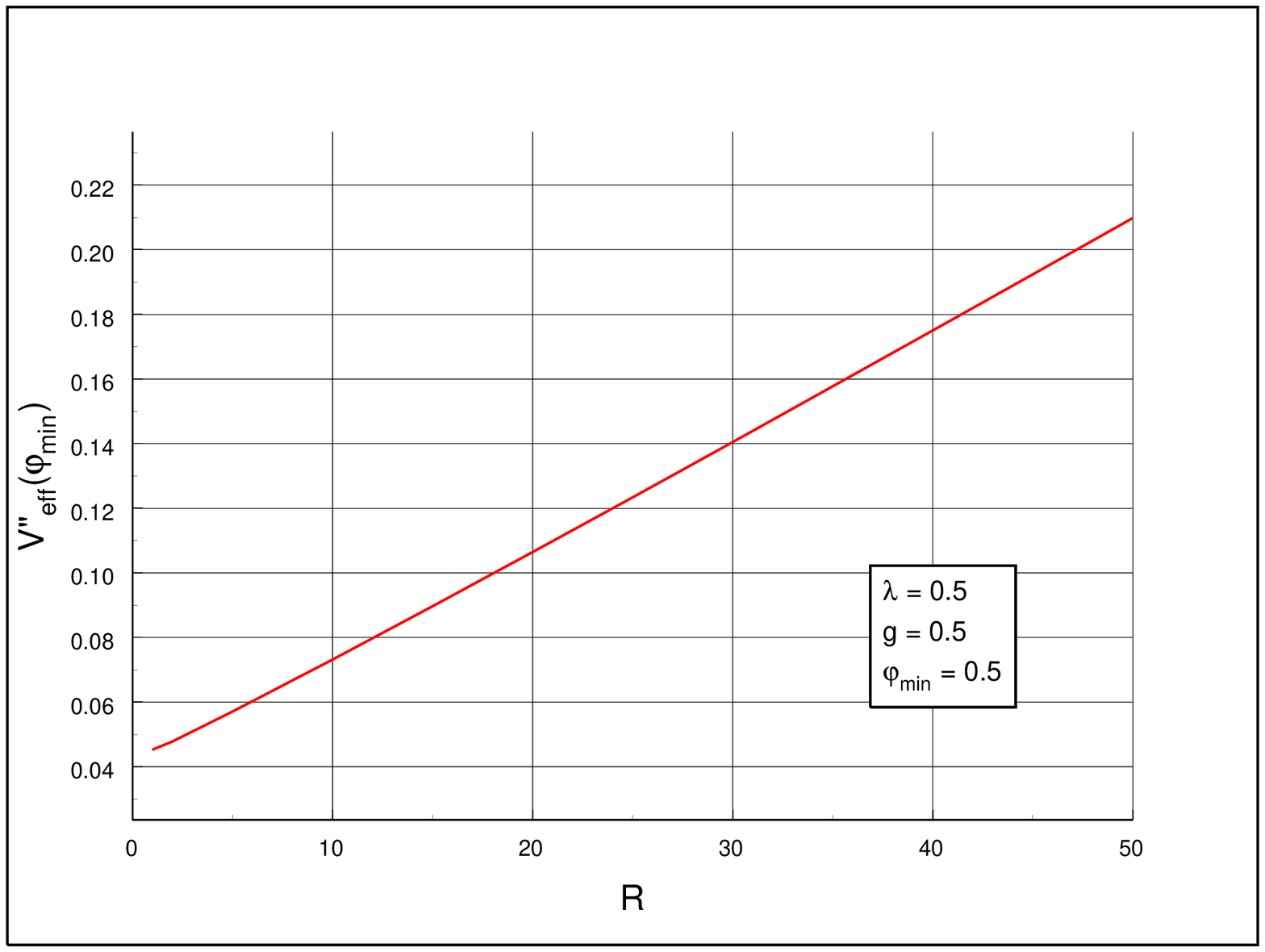 x=6.0in y=3.0in}
\caption{The curvature at the minimum of the effective potential, 
$V_{\rm eff}^{\prime\prime}(\varphi_{\rm min})$ as a function of the size of the compact dimension, $R$.}
\end{figure}

In figure 3, the effective potential is once again plotted  
\begin{figure}[ht]
\vspace*{3.0in}
\special{eps: d: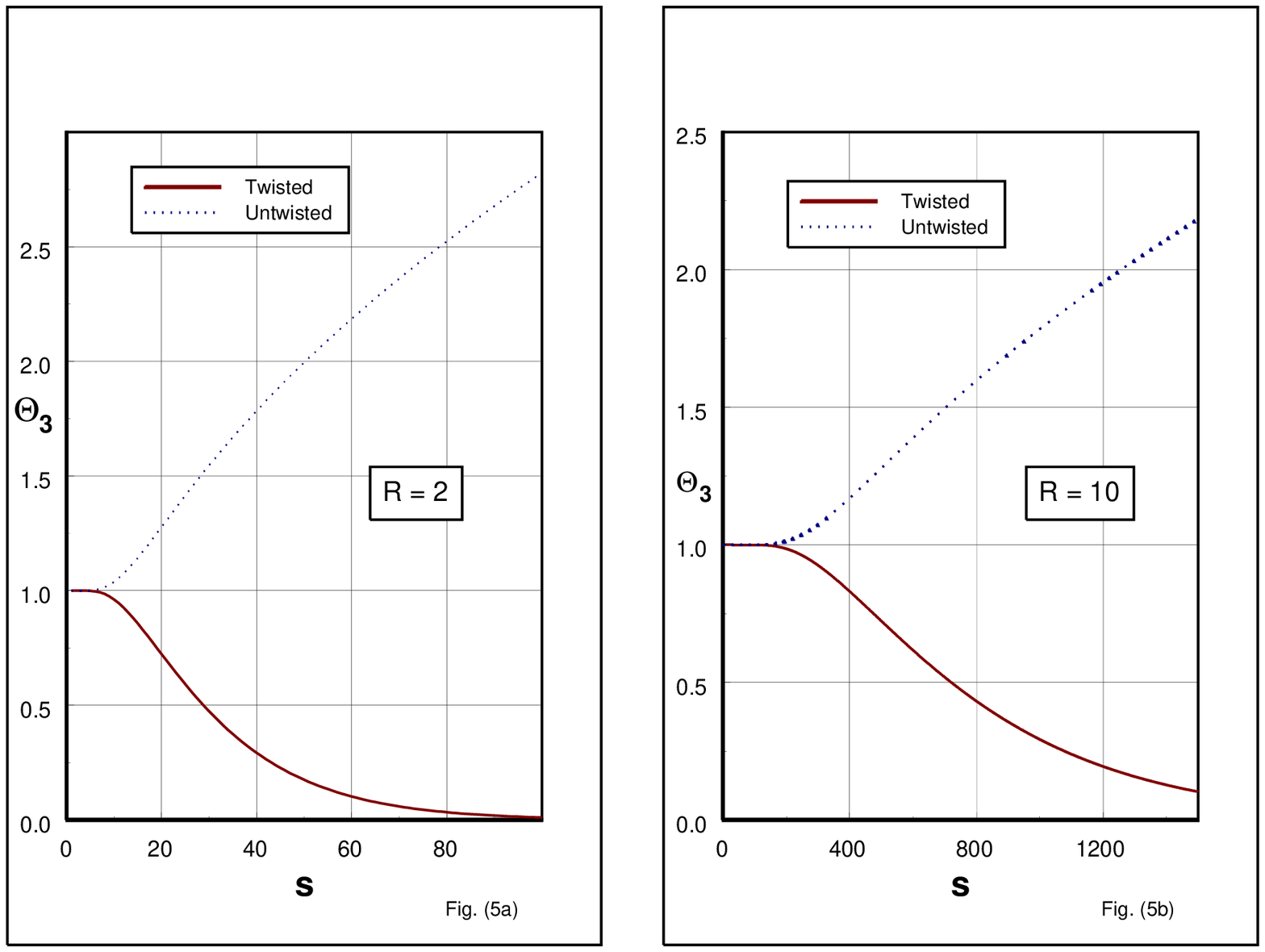 x=6.0in y=3.0in}
\caption{The Jacobi theta function $\Theta_3(\pi a,e^{-\frac{\pi^2 R^2}{s}})$ as a function of $s$ for untwisted ($a=0$) and twisted ($a=\frac{1}{2}$) boundary conditions on the fermions. Note the different abscissa scales in the 2 graphs.}
\end{figure}
as a function of $\varphi$. This time, however, the scale $\mu$ is chosen such that the effective potential vanishes at its minimum which is in turn fixed 
at $\varphi_{\rm min}=0.5=\mu$. In figure (3a) [figure (3b)], it is scalar [Yukawa] coupling which dominates the radiative corrections.  The curvature at the minimum for the same set of parameters as in figure (3b) is displayed in figure 4. It increases essentially linearly with the size of the fifth dimension once again reflecting the explicit $R$ dependence of the radiative corrections. 

Finally, we examine the model where the fermions traversing the loop are taken to have twisted boundary conditions, 
\begin{figure}[ht]
\vspace*{3.0in}
\special{eps: d: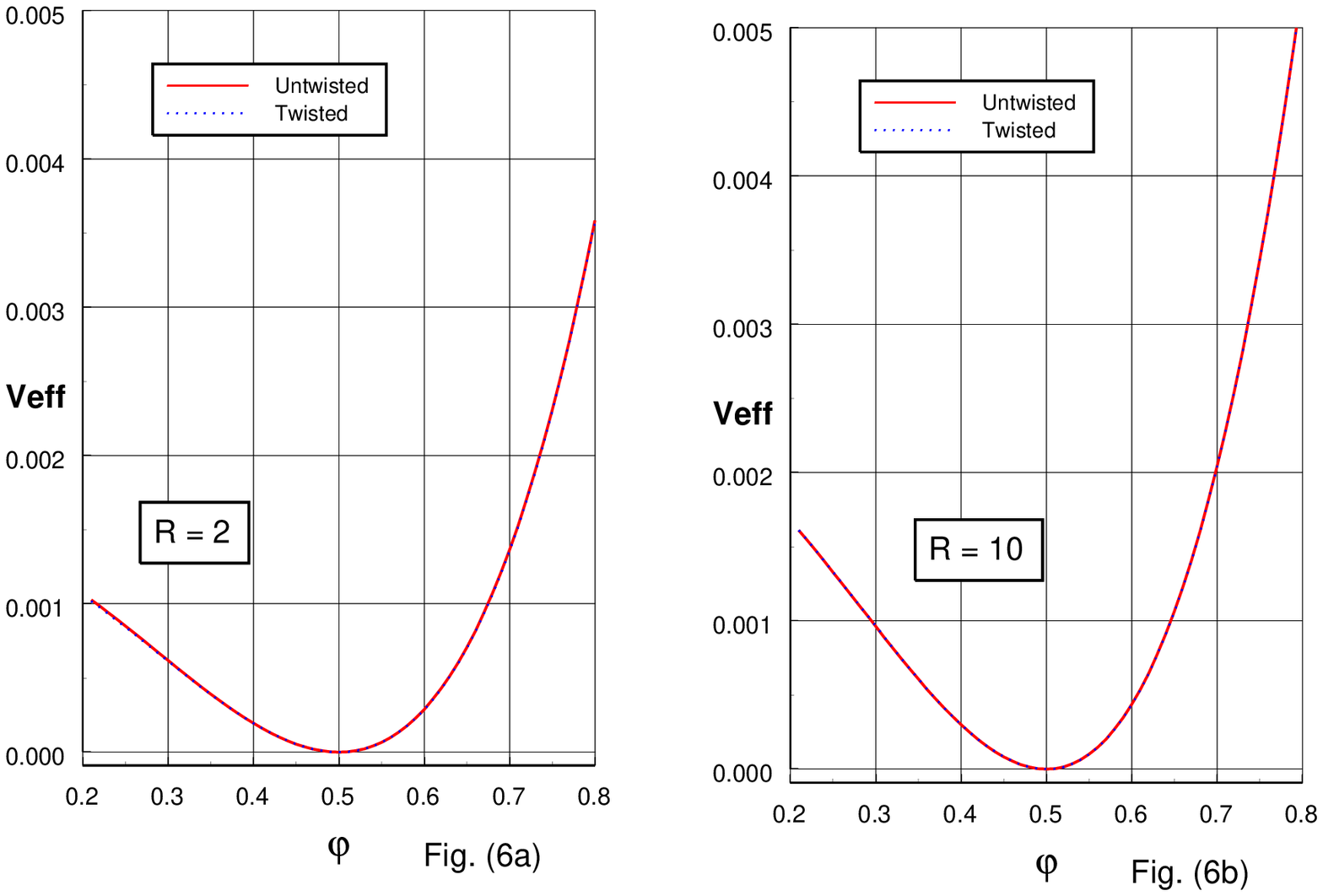 x=6.0in y=3.0in}
\caption{$V_{\rm eff}$ as a function of $\varphi$ for two different compactification radii when the fermions satisfy  both twisted and untwisted boundary conditions. The couplings used are identical to those employed in figure (3b). Although visibly indistinguishable, there are actually 2 curves plotted in each figure.}
\end{figure}
$a= \frac{1}{2}$, on the compact dimensions.  Note that the boundary condition dependence enters the radiative corrections only through the Poisson resummed Jacobi theta function,
\be
\Theta_3 (\pi a, e^{-\frac{\pi^2 R^2}{s}}) = \sum_{n=-\infty}^{\infty}  e^{-\frac{\pi^2 R^2 n^2}{s}} e^{2in\pi a}.
\ee 
which for $a=\frac{1}{2}$ reduces to 
\be
\Theta_3 (\frac{\pi}{2}, e^{-\frac{\pi^2 R^2}{s}}) = 1 + 2\sum_{n=1}^{\infty} (-)^n e^{-\frac{\pi^2 R^2 n^2}{s}}.
\ee 
In figure 5, this function and the analogous function for untwisted boundary condition ($a=0$) are plotted for both $R=2$ and $R=10$. In both cases the function arising from twisted boundary conditions is seen to fall to zero for large $s$. In this limit, the Jacobi theta function is more easily evaluated using its form before applying the Poisson resummmation formula (cf. Eq. {$\ref{PRF}$}). For any twisting in the boundary condition ($a\ne 0$), the function is exponentially surpressed for large $s$. On the other hand, for untwisted boundary conditions ($a=0$) the function grows like $\sqrt{s}$ for large $s$. In the $R=10$ case, the 
two functions are seen to be indistinguishable from unity out to a very large value of the heat kernel parameter $s$ which corresponds to the far infrared. Since for the large $s$ region, the integrand in the heat kernel expression for the effective potential is surpressed by $s^{-7/2}$, both the twisted and untwisted fermions produce essentially identical results as seen in figure 6. 
For the $R=2$ case, the two theta functions deviate from each other at a somewhat smaller value of $s$ where the integrand is not as significantly damped. Nonetheless the explicit numerical integration still gives only a tiny change in the effective potential arising from the different boundary conditions.

In summary, the consequences of large radius extra space-time compactified dimensions on the four dimensional one loop effective potential have been determined for a model which includes scalar self interactions and Yukawa coupling to fermions.  For the case in which the scalar self coupling dominates the radiative corrections, as $R$ increases, the location of the minimum of the effective potential tends to decrease while the potential becomes more shallow.  On the other hand, when the Yukawa coupling dominates the radiative corrections the location of the minimum increases and the minimum deepens as $R$ increases.   With the location of the minimum fixed, the curvature at the minimum increases in the Yukawa coupling dominated case and decreases in the dominant scalar self coupling case as $R$ increases.  Finally the effects of twisted 
boundary conditions for the fermions were compared to those of untwisted boundary conditions.  The twist of the fermion boundary conditions were shown to have but a small effect on the shape of the effective potential \cite{CLVEFF}.
\bigskip

\noindent
This work was supported in part by the U.S. Department 
of Energy under grant DE-FG02-91ER40681 (Task B).

\newpage
\end{document}